\documentstyle[prl,aps,floats,epsf]{revtex}

\begin{document}

\draft

\twocolumn[\hsize\textwidth\columnwidth\hsize\csname @twocolumnfalse\endcsname
\title{Quantum Phase Transition of
  Randomly-Diluted Heisenberg Antiferromagnet \\
  on a Square Lattice}
\author{K.~Kato$^1$\cite{Kato}, S.~Todo$^1$, K.~Harada$^2$, N.~Kawashima$^3$, 
 S.~Miyashita$^{4}$, and H.~Takayama$^1$}
\address{${}^1$Institute for Solid State Physics, University of Tokyo,
  Tokyo 106-8666, Japan}
\address{${}^2$Department of Applied Analysis and Complex Dynamical Systems,
Kyoto University, Kyoto 606-8501, Japan}
\address{${}^3$Department of Physics, Tokyo Metropolitan University,
  Tokyo 192-0397, Japan}
\address{${}^4$Department of Applied Physics, University of Tokyo,
  Tokyo 113-8656, Japan}
\date{\today}
\maketitle

\widetext

\begin{abstract}
 Ground-state magnetic properties of the diluted Heisenberg
 antiferromagnet on a square lattice are investigated by means of the
 quantum Monte Carlo method with the continuous-time loop algorithm.  It
 is found that the critical concentration of magnetic sites is
 independent of the spin size $S$, and equal to the two-dimensional
 percolation threshold.  However, the existence of quantum fluctuations
 makes the critical exponents deviate from those of the classical percolation
 transition.  Furthermore, we found that the transition is not
 universal, i.e., the critical exponents significantly depend on $S$.
\end{abstract}
]
\narrowtext

 Since the discovery of high-temperature superconductivity in
 cuprates~\cite{Bednorz86}, effects of non-magnetic impurities in
 two-dimensional Heisenberg antiferromagnets (HAF's) have been studied
 extensively.  As is well known, the antiferromagnetic (AF) long-range
 order in pure La$_2$CuO$_4$ ($S=1/2$ HAF) is immediately destroyed by a
 small amount of Sr substitution for La.  {\em Itinerant} holes, doped
 into the CuO$_2$ plane, cause magnetic frustration among Cu $S=1/2$
 spins and destroy the AF order.  On the other hand, the AF long-range
 order is much robust~\cite{Jongh83,Johnston87,Cheong91,Corti95} against
 {\em static} magnetic impurities such as $S=0$ Mg or Zn in
 La$_2$CuO$_4$, and Mg in K$_2$CoF$_4$ ($S=1/2$ Ising AF) or
 K$_2$MnF$_4$ (S=5/2 HAF).  However, it is notable that the critical
 concentration of non-magnetic impurities strongly depends on the system
 in these cases.  The critical concentration of
 La$_2$Cu$_{1-x}$Mg$_x$O$_4$ and La$_2$Cu$_{1-x}$Zn$_x$O$_4$ is
 observed~\cite{Cheong91} as $x\sim20\%$.  It makes a sharp contrast
 with the others~\cite{Jongh83,Johnston87}, where the critical
 concentration is almost equal to the two-dimensional percolation
 threshold ($x\sim40\%$).  In La$_2$Cu$_{1-x}$Mg$_x$O$_4$ and
 La$_2$Cu$_{1-x}$Zn$_x$O$_4$, a {\em quantum disordered phase} may be
 realized due to strong quantum fluctuations.

 These experimental results, and also purely theoretical interests, have
 motivated recent theoretical and also numerical
 studies~\cite{MiyashitaBY94,YasudaO97,ChenN99} on the site-diluted HAF
 on the square lattice:
 \begin{equation}
  \label{Hamiltonian}
   {\cal H} = J \sum_{\langle i,j \rangle} \epsilon_i\,\epsilon_j \,
   {\bf S}_i \cdot {\bf S}_j \,.
 \end{equation}
 Here, ${\bf S}_i = (S_i^x,S_i^y,S_i^z)$ denotes the the quantum
 spin-$S$ operator at site $i$, and the nearest-neighbor coupling
 constant is antiferromagnetic ($J>0$).  The quenched dilution factors
 $\{\epsilon_i\}$ independently take 1 or 0 with probability $p$ and
 $1-p$, respectively, where $p$ ($=1-x$) denotes the concentration of
 magnetic sites.

 In the classical case ($S=\infty$), the present model at zero
 temperature is equivalent to the site-percolation
 problem~\cite{Percolation}.  It is well known that the system undergoes
 a second-order phase transition at the percolation threshold $p_{\rm
 cl}$, which is determined as
 \begin{equation}
  p_{\rm cl} = 0.5927460(5)
 \end{equation}
 by the most recent simulation~\cite{Ziff92}.  Near $p_{\rm cl}$, the
 staggered magnetization vanishes as $M_{\rm s} \sim (p-p_{\rm
 cl})^{\beta_{\rm cl}}$ with $\beta_{\rm cl} = 5/36 =
 0.13888\cdots$~\cite{Percolation}.

 The main subjects in the previous
 works~\cite{MiyashitaBY94,YasudaO97,ChenN99} have been focused on
 whether the critical concentration of the quantum systems ($S<\infty$),
 referred to as $p^*$ hereafter, is identical to that of the classical
 model, $p_{\rm cl}$, or not, and also on its dependence on the strength
 of quantum fluctuations specified by the spin size $S$.  For $S=1/2$,
 $p^* > p_{\rm cl}$ is suggested in these
 studies~\cite{MiyashitaBY94,YasudaO97,ChenN99,SandvikV95}.  For
 example, $p^*=0.655$ and 0.695 are obtained in \cite{MiyashitaBY94} and
 \cite{ChenN99}, respectively.  However, they are still inconclusive
 and critical properties of the quantum phase transition and the quantum
 disordered phase (if exists) between $p^*$ and $p_{\rm cl}$ have not
 been well understood.

 In this paper, we report results of large-scale quantum Monte Carlo
 (QMC) simulations on the diluted HAF (\ref{Hamiltonian}) with $S=1/2$,
 1, $3/2$, and 2.  It is found, as we see below, that the AF long-range
 order at $T=0$ persists so long as a cluster of magnetic sites
 percolates, that is, $p^*=p_{\rm cl}$, even in the $S=1/2$ case.
 However, we find the critical exponents of the quantum phase transition
 just at $p_{\rm cl}$ are clearly different from those of the classical
 transition; the critical exponents vary depending on the value of $S$.

 In the present QMC simulation, use of the continuous-time loop
 algorithm~\cite{Evertz97,Loop,HaradaTK97,TodoK99} is crucial.  It
 greatly reduces correlations between successive world-line
 configurations, and therefore makes it possible to perform highly
 reliable simulations on large lattices (up to $L \times L= 48\times48$)
 at extremely low temperatures ($T\sim0.001J$).  Another important
 feature of the present algorithm is its ergodicity; the winding number
 of world lines around vacant sites can change and the ground-canonical
 ensemble can be simulated.  We use the periodic boundary conditions.
 For each sample, $10^4$ Monte Carlo steps (MCS) are spent for
 measurement after $10^3$ MCS for thermalization.  At each parameter set
 $(L,T,p)$, physical quantities are averaged over 100 -- 1000 samples.

 \begin{figure}[tb]
  \hspace*{0em}\epsfxsize=0.47\textwidth
  \epsfbox{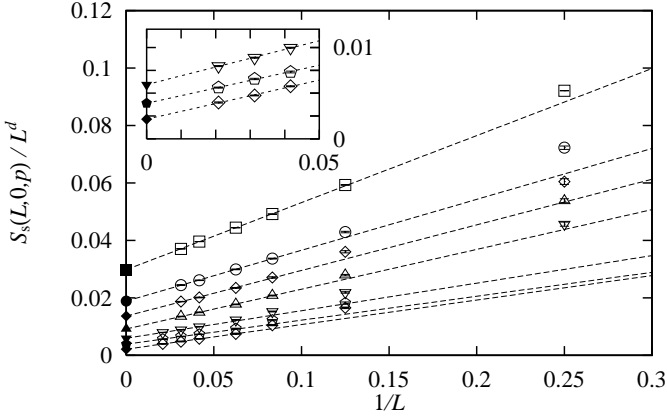}
  \caption{System-size dependence of $S_{\rm s}(L,T=0,p)/L^d$ in the case
  of $S=1/2$ at $p=1$, 0.875, 0.8125, 0.75, 0.6875, 0.65, and 0.625 (from
  upper to lower).  The dashed lines are obtained by least-squares
  fitting for the largest three system sizes for each $p$.  The
  extrapolated values are denoted by solid symbols.  The data with
  $p\le0.6875$ and $L\ge24$ are also shown in the inset.}
  \label{fig:ss_of_l}
 \end{figure}

 \begin{figure}[tb]
  \epsfxsize=0.47\textwidth
  \epsfbox{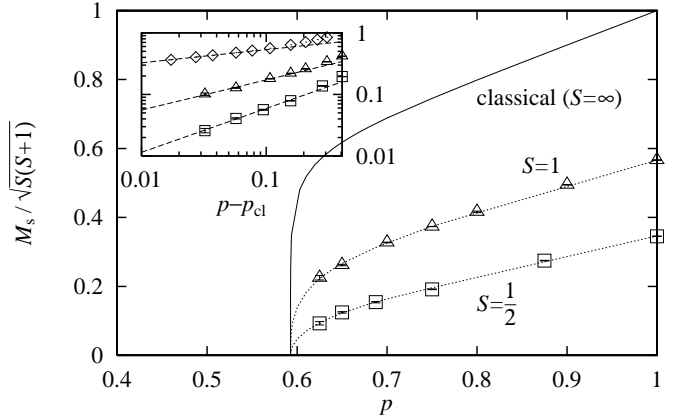}
  \caption{Concentration dependence of the (normalized) staggered
  magnetization in the cases of $S=1/2$ (square) and 1 (triangle).  The
  dotted lines are guides to eyes.  In the inset, we show the
  double-logarithmic plot of the staggered magnetization against
  $(p-p_{\rm cl})$.  The dashed lines are obtained by least-squares
  fitting for $p\le0.70$.  For comparison, the staggered magnetization
  in the classical limit is also plotted by the solid line and
  diamonds.}
  \label{fig:ms_of_p}
 \end{figure}

 First, we discuss the zero-temperature staggered magnetization $M_{\rm
 s}(p)$ for $p>p_{\rm cl}$, which is calculated as
 \begin{eqnarray}
  \label{staggered-magnetization}
   M_{\rm s}^2(p) = \lim_{L \to \infty} \lim_{T \to 0}
   \frac{3 S_{\rm s}(L,T,p)}{L^d}
 \end{eqnarray}
 in terms of the static structure factor defined by
 \begin{equation}
  \label{structure-factor}
   S_{\rm s}(L,T,p) = \frac{1}{L^d} \sum_{i,j} e^{i \vec{k} \cdot
   (\vec{r}_i - \vec{r}_j)} \langle S^z_i S^z_j \rangle
 \end{equation}
 at the momentum $\vec{k}=(\pi,\pi)$, where $d$ is the spatial dimension
 ($d=2$).  The bracket in Eq.~(\ref{structure-factor}) denotes both of
 the thermal average and the average over samples.  In the present
 simulation, we use an improved estimator to calculate $S_{\rm
 s}(L,T,p)$, by which the variance of data is greatly reduced.
 
 At temperatures lower than the gap of a finite system, $S_{\rm
 s}(L,T,p)$ converges to its zero-temperature value quite rapidly
 (probably exponentially).  To obtain $S_{\rm s}(L,T=0,p)$ at each $p$
 and $L$, we perform QMC simulations at low enough temperatures so that
 $S_{\rm s}(L,T,p)$ exhibits no temperature dependence besides
 statistical errors.  Note that as $p$ decreases, the gap of a system of
 linear size $L$ becomes smaller, and therefore lower temperature is
 needed~\cite{fn-2}.  For the $S=1/2$ case, $T$ is taken as $0.002J$ at
 $p=0.625$ and $L=48$.

 In Fig.~\ref{fig:ss_of_l}, we plot $S_{\rm s}(L,0,p)/L^d$ against $1/L$
 for $0.625 \le p \le 1$ in the $S=1/2$ case.  It is clearly seen that
 the data at each concentration fall on a straight line for large $L$.
 In the clean system ($p=1$), the leading finite-size correction is
 shown to be of $O(1/L)$ according to the spin-wave
 theory~\cite{Huse88}.  The similar behavior for $p<1$ indicates that
 there exists an AF long-range order with massless excitations, such as
 spin waves, even in the presence of impurities.  Thus, the staggered
 magnetization in the thermodynamic limit is obtained by linear
 extrapolation in $1/L$ for three largest system sizes at each $p$.
 
\begin{figure}[tb]
  \epsfxsize=0.47\textwidth
  \epsfbox{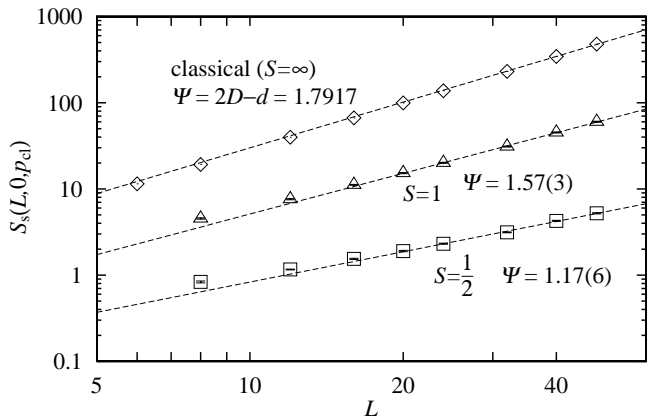}
  \caption{System-size dependence of $S_{\rm s}(L,T=0,p)$ at $p=p_{\rm
  cl}$ in the cases of $S=1/2$ (square) and 1 (triangle).  The dashed
  lines are obtained by least-squares fitting for $L\ge24$.  The slope of
  the lines gives the exponent $\Psi$ (Eq.~(\protect\ref{eqn:scale_ss-0})).
  The data in the classical limit are also plotted by diamonds.}
  \label{fig:ss_at_pc}
 \end{figure}

 The final results for the zero-temperature staggered magnetization are
 shown in Fig.~\ref{fig:ms_of_p}.  It is seen that the staggered
 magnetization remains finite even at $p=0.625$ both in the $S=1/2$ and
 1 cases.  The possibility that $p^*$ is greater than $0.625$ is
 excluded definitely, and the behavior of the whole magnetization curve
 strongly suggests that the critical concentration is identical to the
 percolation threshold ($p^*=p_{\rm cl}$).  Actually, the staggered
 magnetization seems to vanish algebraically towards $p=p_{\rm cl}$ as
 seen in the inset of Fig~\ref{fig:ms_of_p}.  By least-squares fitting,
 we estimate the critical exponent $\beta$ as 0.46(3) and 0.32(3) for
 $S=1/2$ and 1, respectively.  Our conjecture, $p^*=p_{\rm cl}$, is also
 supported by the power-law behavior of the zero-temperature static
 structure factor just at $p=p_{\rm cl}$:
 \begin{equation}
  \label{eqn:scale_ss-0}
  S_{\rm s}(L,0,p_{\rm cl}) \sim L^\Psi
 \end{equation}
 with $\Psi=1.17(6)$ and 1.57(3) for $S=1/2$ and 1, respectively
 (Fig.~\ref{fig:ss_at_pc}).
 
 The critical exponents $\beta$ and $\Psi$, obtained by the above
 analysis, differ from those of the percolation model ($\beta=5/36$ and
 $\Psi=43/24$~\cite{Percolation}), that is, the universality class of
 the quantum phase transition at $p=p_{\rm cl}$ is different from the
 classical one.  Furthermore, we found that the value of the critical
 exponents depends on $S$.  This means that not only the fractal nature
 of the lattice geometry, but also the existence of quantum fluctuations
 and their strength, or the value of $S$, are relevant to the
 criticality.  Hence, it is natural to introduce the two following
 assumptions on the spin correlation on spin clusters.
 
 \begin{table}[b]
  \caption{Summary of critical exponents $\Psi$, $z$, $\beta$, and $\nu$.
  $\Psi$ and $z$ are obtained by the FSS shown in
  Fig.~\protect\ref{fig:sc_plot}, and $\beta$ is estimated from the
  staggered magnetization (Fig.~\protect\ref{fig:ms_of_p}).  The exponent
  $\nu$ is calculated from $\beta$ and $\Psi$ thorough the scaling
  relations, Eqs.~(\protect\ref{eqn:scale_rel-1}) and
  (\protect\ref{eqn:scale_rel-2}).}
  \label{tab:exponents}
  \begin{tabular}{cllll}
   $S$ & \ \ \ $\Psi$ & \ \ \ $z$ & \ \ $\beta$ & \ \ $\nu$ \\
   \hline
   $1/2$    & 1.27(2) & 2.54(8)  & 0.46(3)   & 1.2(1) \\
   $1$      & 1.57(3) & 1.58(10) & 0.32(3)   & 1.5(2) \\
   $3/2$    & 1.60(3) & 1.55(10) & \ \ \ --- & \ \ \ --- \\
   $2$      & 1.69(7) & 1.31(20) & \ \ \ --- & \ \ \ --- \\
   \hline
   $\infty$ & 1.79167 & \ \ \ ---& 0.13889   & 1.33333 \\
  \end{tabular}
 \end{table}
 
 At $p=p_{\rm cl}$, all clusters are fractal with a fractal dimension
 $D$ (in two dimensions, $D=91/48$)~\cite{Percolation}.  First, we
 assume that the staggered spin correlation between two sites, $i$ and
 $j$, on a cluster behaves as
 \begin{equation}
  \label{eqn:corr-1}
  C(i,j) \sim r_{i,j}^{-\alpha} \ \ \ 
   \text{for $r_{i,j} \gg 1$,}
 \end{equation}
 where $r_{i,j}=|\, \vec{r_i}-\vec{r_j} \,|$.  Here, we introduce a new
 $S$-dependent exponent $\alpha=\alpha(S)\ge0$.  In the classical case,
 $C(i,j)$ takes a constant independent of $r_{i,j}$, i.e.,
 $\alpha(\infty)=0$.  Together with the cluster-size distribution at
 $p=p_{\rm cl}$, predicted by the percolation theory~\cite{Percolation},
 we obtain
 \begin{equation}
  \label{eqn:scale_ss-1} S_{\rm s} (L,0,p_{\rm cl}) = \frac{1}{L^d}
  \sum_{\Omega} \sum_{i,j \in \Omega} C(i,j) \sim
  L^{2D-d-\alpha} \,,
 \end{equation}
 where the first summation on the first line is taken over clusters on
 the lattice.  Comparing Eq.~(\ref{eqn:scale_ss-1}) with
 Eq.~(\ref{eqn:scale_ss-0}), we obtain a scaling relation:
 \begin{equation}
  \label{eqn:scale_rel-1}
  \Psi = 2D-d-\alpha \,.
 \end{equation}

 On the other hand, for $p>p_{\rm cl}$, the percolation
 theory~\cite{Percolation} says that there exists a characteristic
 length $\lambda(p)$, below which the percolated cluster exhibits a
 fractal nature similar to that of clusters at $p=p_{\rm cl}$, but it has
 the ordinary dimension ($d=2$) at longer length scale.  For
 $L\gg\lambda(p)$, $S_{\rm s}(L,T,p)$ is dominated by the percolated
 cluster.  We introduce the second assumption that there exist {\em no}
 other macroscopic characteristic lengths except $\lambda(p)$ even in
 the $S<\infty$ cases.  Since there exists an AF long-range order as
 shown before, the correlation function is expected to obey the
 following scaling form~\cite{fn-1}:
 \begin{eqnarray}
  \label{eqn:corr-2}
  C(i,j;p) 
  &\sim& r_{i,j}^{-\alpha} \tilde{C}(r_{i,j}/\lambda(p)) \nonumber \\
  &\sim& \lambda(p)^{-\alpha} \ \ \ 
   \text{for $r_{i,j} \gg \lambda(p)$} \,,
 \end{eqnarray}
 where $\tilde{C}(x)={\rm const.}$ for $x\rightarrow0$, so that
 Eq.~(\ref{eqn:corr-1}) is reproduced for $1 \ll r_{i,j} \ll
 \lambda(p)$.  

\begin{figure}[tb]
  \epsfxsize=0.47\textwidth
  \epsfbox{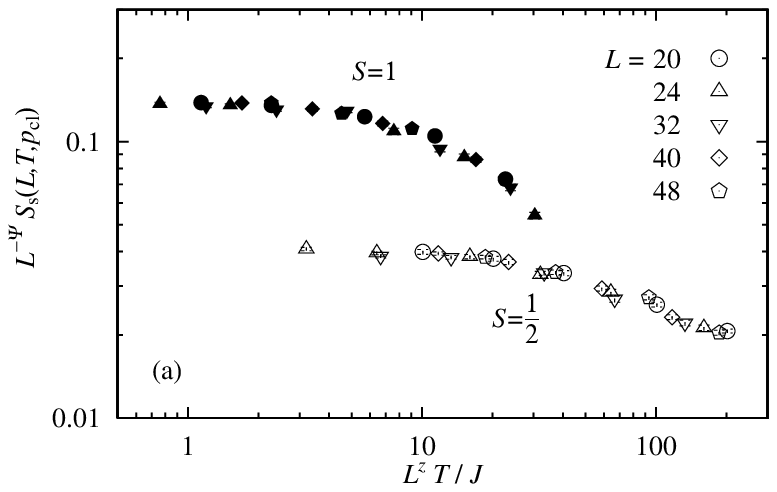}
  
  \vspace*{-.7em}\hspace*{.6em}
  \epsfxsize=0.457\textwidth
  \epsfbox{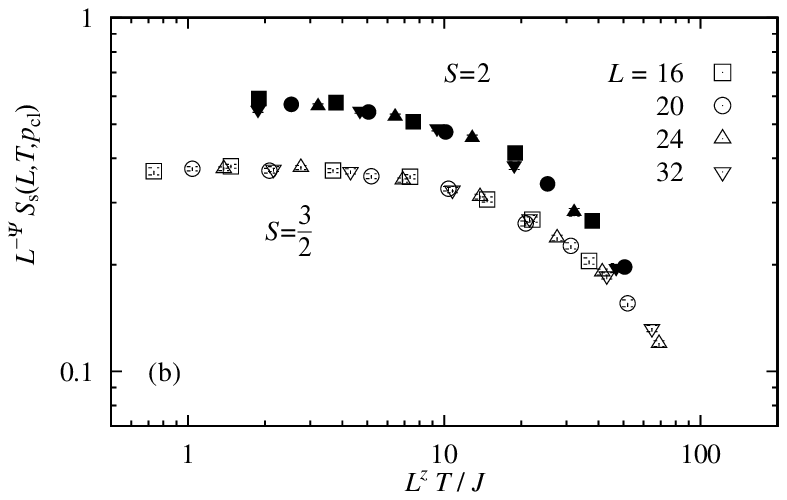}
  \caption{Scaling plot of $S_{\rm s}(L,T,p)$ at $p=p_{\rm cl}$ for (a)
  $S=1/2$ and 1, and for (b) $S=3/2$ and 2.}
  \label{fig:sc_plot}
 \end{figure}

 Using the fact that the characteristic length scale $\lambda(p)$
 diverges as $(p-p_{\rm cl})^{-\nu}$ near $p_{\rm cl}$ with
 $\nu=4/3$~\cite{Percolation}, and assuming the ordinary algebraic
 temperature dependence, one finally reaches a full finite-size scaling
 (FSS) form of the structure factor:
 \begin{eqnarray} 
  \label{scale-ss0}
  S_{\rm s}(L,T,p)
  & \sim & L^{2D-d-\alpha} \tilde{S}_{\rm s}
  (L^{1/\nu} (p-p_{\rm cl}),L^z T) \,,
 \end{eqnarray}
 where the $S$-dependent scaling function $\tilde{S}_{\rm s}(x,y)$ takes
 a constant at $(x,y)=(0,0)$, and $\tilde{S}_{\rm s}(x,0) \sim
 x^{-(2D-2d-\alpha)\nu}$ for $x\gg1$.  In Eq.~(\ref{scale-ss0}), we
 introduce the dynamical exponent $z$, which relates the energy scale
 with the length scale~\cite{Hertz76}.  In the present case, $z>1$ is
 expected, since the Lorentz invariance is broken due to the existence
 of impurities.  Although the algebraic $T$ dependence assumed in
 Eq.~(\ref{scale-ss0}) is not guaranteed {\em a priori}, finite-temperature
 data of the staggered structure factor at $p=p_{\rm cl}$ is well scaled
 for $S=1/2$, 1, 3/2, and 2, by using our FSS form (\ref{scale-ss0})
 with the exponents $\Psi$ and $z$ listed in TABLE~\ref{tab:exponents},
 as shown in Fig.~\ref{fig:sc_plot}.
 The values of $\Psi$ are consistent with those obtained by the FSS at
 $T=0$ (Eq.~(\ref{eqn:scale_ss-0}) and Fig.~\ref{fig:ss_at_pc}).  Note
 that not only $\Psi$ but also $z$ depend on $S$.  In addition, $\Psi$
 seems to approach the classical value monotonically as $S$ increases.

 The critical exponent $\beta$ is also expressed in terms of $\alpha$.
 By taking $L\rightarrow\infty$ after taking the limit of
 $T\rightarrow0$ in Eq.~(\ref{scale-ss0}), one obtains
 \begin{equation}
  \label{eqn:scale_rel-2}
  2\beta = -(2D-2d-\alpha)\nu \,.
 \end{equation}
 The values of $\nu$ calculated from $\Psi$ and $\beta$ through
 Eqs.~(\ref{eqn:scale_rel-1}) and (\ref{eqn:scale_rel-2}) are satisfactorily
 consistent with $\nu=4/3$ (TABLE~\ref{tab:exponents}), which supports
 the validity of the scaling argument discussed above.

 In summary, we have investigated the ground-state phase transition of
 the diluted HAF with $S=1/2$, 1, 3/2, and 2.  Contrary to the previous
 works~\cite{MiyashitaBY94,YasudaO97,ChenN99,SandvikV95}, our present
 QMC study has shown that the critical concentration is equal to the
 classical percolation threshold even in the $S=1/2$ case.  Concerning
 the relation to experiments, it becomes clear that the present model
 (\ref{Hamiltonian}) is too much idealized to predict the magnetic
 properties observed in the real materials.  Other effects, such as
 next-nearest-neighbor interaction, should be included properly into the
 model Hamiltonian.  On the other hand, in the theoretical point of
 view, it has been revealed that the present model contains quite rich
 physics; the critical exponents vary depending on $S$.  We have
 introduced a scaling form with two non-classical exponents, $\alpha$
 and $z$, which consistently explains the behavior of the simulation
 data together with the exponents $D$ and $\nu$ of the classical
 percolation transition.  To our best knowledge, this is the first time
 that such an $S$-dependent quantum critical behavior is discovered.
 
 Most of numerical calculations for the present work have been performed
 on the CP-PACS at University of Tsukuba, Hitachi SR-2201 at
 Supercomputer Center, University of Tokyo, and on the RANDOM at
 Materials Design and Characterization Laboratory, Institute for Solid
 State Physics, University of Tokyo.  The present work is supported by
 the ``Large-scale Numerical Simulation Program'' of Center for
 Computational Physics, University of Tsukuba, and also by the
 ``Research for the Future Program (Computational Science and
 Engineering)'' of Japan Society for the Promotion of Science.  N.K.'s
 work is supported by Grant-in-Aid for Scientific Research Program
 (No.09740320) from the Ministry of Education, Science, Sport and
 Culture of Japan.


\begin{thebibliography}{99}
\bibitem{fn-2} The temperatures ($T\sim0.1J$) used in the previous QMC
	     studies~\cite{MiyashitaBY94,SandvikV95} are too high to
	     estimate correct zero-temperature staggered magnetization
	     for small $p$.
\bibitem{fn-1} We omitted corrections from massless excitations in
	     Eq.~(\ref{eqn:corr-2}) for simplicity.
\end{thebibliography}
\end{document}